\documentclass[9pt,twocolumn,twoside]{osajnl}

\journal{ol}

\setboolean{shortarticle}{true} 

\ifthenelse{\boolean{shortarticle}}{\colorlet{color2}{color2b}}{\colorlet{color2}{color2}}

\title{Experimental observation of three-photon superbunching in linear optical system}
\author[1]{Yu Zhou}
\author[2]{Sheng Luo}
\author[2]{Zhaohui Tang}
\author[2]{Huaibin Zheng}
\author[2]{Hui Chen}
\author[2,*]{Jianbin Liu}
\author[1]{Fu-li Li}
\author[2]{Zhuo Xu}

\affil[1]{MOE Key Laboratory for Nonequilibrium Synthesis and Modulation of Condensed Matter, Department of Applied Physics, Xi'an Jiaotong University, Xi'an 710049, China}
\affil[2]{Electronic Materials Research Laboratory, Key Laboratory of the Ministry of Education \& International Center for Dielectric Research, School of Electronic and Information Engineering, Xi'an Jiaotong University, Xi'an 710049, China}

\affil[*]{Corresponding author: liujianbin@xjtu.edu.cn}

\dates{Compiled \today}

\ociscodes{(270.5290) Photon statistics; (030.1640) Coherence.}

\doi{\url{http://dx.doi.org/10.1364/OL.XX.XXXXXX}}

\begin{abstract}
Three-photon superbunching is experimentally observed with the recently proposed superbunching pseudothermal light. To the best of our knowledge, it is the first time that three-photon superbunching is observed in linear optical system. From the quantum optics point of view, three-photon superbunching is interpreted as the result of constructive-destructive three-photon interference. The key to observe three-photon superbunching in superbunching pseudothermal light is that all the different ways to trigger a three-photon coincidence count are in principle indistinguishable, which is experimentally guaranteed by putting a pinhole before each rotating groundglass to ensure that all the passed photons are within one coherence area. The observed three-photon superbunching is helpful to increase the visibility of ghost imaging and understand the physics of third-order interference of light.
\end{abstract}

\setboolean{displaycopyright}{true}

\begin{document}

\maketitle
\thispagestyle{fancy}

\ifthenelse{\boolean{shortarticle}}{\ifthenelse{\boolean{singlecolumn}}{\abscontentformatted}{\abscontent}}{}

Photon bunching of thermal light, first discovered by Hanbury Brown and Twiss \cite{HBT, HBT1}, is the foundation for ghost imaging with thermal light \cite{wang}. Comparing to ghost imaging with entangled photon pairs \cite{pittman}, ghost imaging with thermal light has been widely applied in different imaging schemes such as imaging through atmospheric turbulence \cite{at-1,at-2,at-3},  imaging through scattering media \cite{sm-1,sm-2,sm-3}, lensless imaging \cite{li-1,li-2} for its simplicity in practical experiments. Similar idea has been extended to ghost imaging with incoherent X-ray \cite{x-1,x-2,x-3} and cold atoms \cite{ca}.  However, one of the drawbacks of thermal light ghost imaging is its low visibility due to the degree of second-order coherence of thermal light is much less than the one of entangled photon pairs \cite{gatti-2004}. Higher-order correlation of thermal light is employed to increase the visibility of thermal light ghost imaging \cite{cao,zhou-2010,chen-2010} since the degree of $N$th-order coherence of thermal light equals $N!$ \cite{liu-2009}, which, however, is not large enough and the complexity of the experimental setup increases as $N$ increases. 

Recently, we have proposed a new type of light called superbunching pseudothermal light \cite{zhou-2017}, which may solve the contradiction between the high visibility and complicated experimental setup in ghost imaging with higher-order coherence of thermal light. The degree of $N$th-order coherence of superbunching pseudothermal light can be increased to $(N!)^n$, where $n$ is the number of rotating groundglass in superbunching pseudothermal light source \cite{zhou-2017}. Most of the existed studies of superbunching pseudothermal light are about the second-order coherence \cite{zhou-2017,bai-2017,liu-2018}. There is only one theoretical study about the third-order coherence of superbunching pseudothermal light \cite{bai-2017}. In this letter, we will experimentally study the third-order coherence of superbunching pseudothermal light. Three-photon superbunching in linear optical system is firstly observed. The degree of third-order coherence equaling 22.55 is obtained by employing two rotating groudglasses, which can further be increased by employing more rotating groundglasses. This new type of light is helpful to increase the visibility of ghost imaging with thermal light, which is also helpful to understand the physics of second-, third-, and higher-order interference of light.

\begin{figure}[htbp]
\centering
\includegraphics[width=80mm]{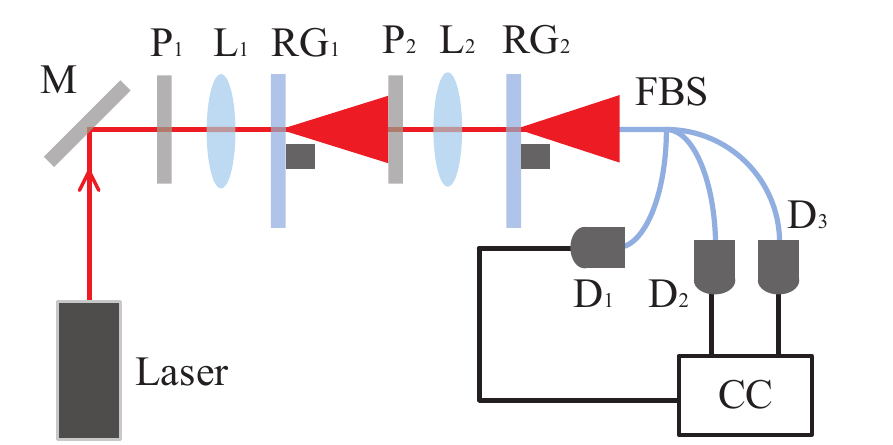}
\caption{Experimental setup to measure three-photon superbunching. Laser: single-mode continuous-wave laser. M: mirror. P: pinhole. L: lens. RG: rotating groundglass. FBS: 1:1:1 fiber beam splitter. D: single-photon detector. CC: three-photon coincidence counting system.}\label{setup}
\end{figure}

The experimental setup to measure three-photon superbunching is shown in Fig. \ref{setup}. Laser is a single-mode continuous-wave laser with central wavelength 780 nm and frequency bandwidth 200 kHz. M is a mirror. P$_1$ and P$_2$ are two pinholes, which are employed to filter the light within one spatial coherence area before RG$_1$ and RG$_2$, respectively. RG is short for rotating groundglass. L$_1$ and L$_2$ are two focus lens to control the size of light spot on RG$_1$ and RG$_2$, respectively. FBS is a 1:1:1 fiber beam splitter. D$_1$,  D$_2$, and  D$_3$ are three single-photon detectors. CC is three-photon coincidence count detection system. The third-order temporal coherence function of superbunching thermal light is measured with the help of FBS, D$_1$, D$_2$, D$_3$, and CC.

\begin{figure}[htbp]
\centering
\includegraphics[width=90mm]{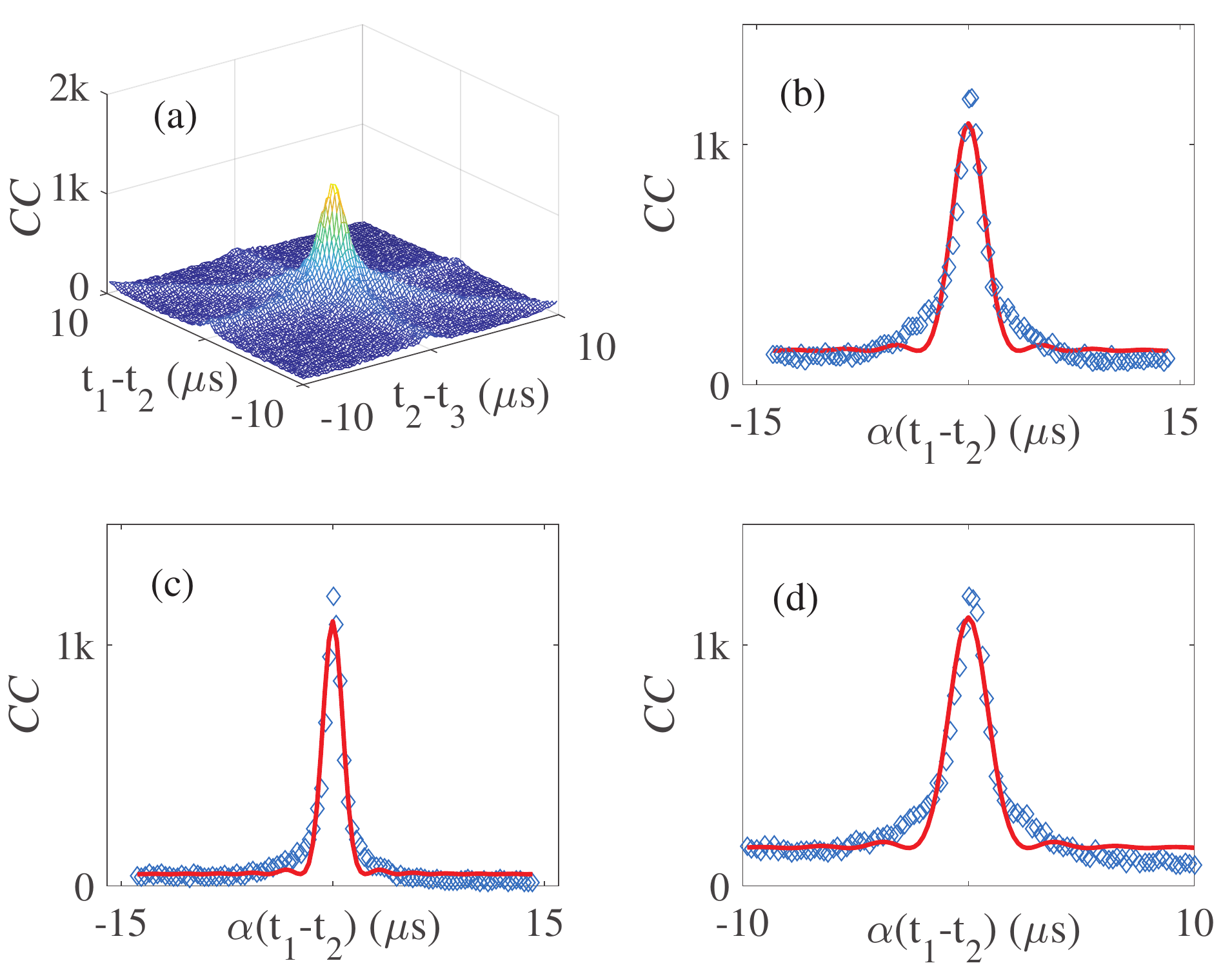}
\caption{Measure three-photon coincidence counts. $CC$: three-photon coincidence counts. $t_1-t_2$: the time difference between photon detection events at D$_1$ and D$_2$ within a three-photon coincidence count. The meaning of $t_2-t_3$ is the same as the one of $t_1-t_2$. $\alpha$ equals $\sqrt{2}$ in (b) and (c), and 1 in (d), which is determined by the slice direction in (a).}\label{super}
\end{figure}

Figure \ref{super}(a) shows the measured three-photon coincidence counts in the experimental setup shown in Fig. \ref{setup}. $CC$ is three-photon coincidence counts. $t_1-t_2$ is the time difference between the photon detection events at D$_1$ and D$_2$ within a three-photon coincidence count. The meaning of $t_2-t_3$ is similar as the one of $t_1-t_2$.  The diamonds and red curves in Figs. \ref{super}(b) - \ref{super}(d) are measured results and theoretical fittings, respectively. $\alpha$ equals $\sqrt{2}$ in Figs. \ref{super}(b) and \ref{super}(c) and equals 1 in Fig. \ref{super}(d), which is determined by the slice direction in Fig. \ref{super}(a). Figure \ref{super}(b) is the section of Fig. \ref{super}(a) along the direction of $t_1-t_2=-(t_2-t_3)$, which corresponds to $t_1$ equaling $t_3$. Figure \ref{super}(d) is the section of Fig. \ref{super}(a) along the direction of $t_1-t_2=0$, which corresponds to $t_1$ equaling $t_2$. Figure \ref{super}(c) is the section of Fig. \ref{super}(a) along the direction of $t_1-t_2=t_2-t_3$, which is employed to calculate the measured degree of third-order coherence of superbunching pseudothermal light. The measured degree of third-order coherence equals 22.55, which is calculated by the fitting the experimental data in Fig. \ref{super}(c) with Eq. (\ref{G2-32}). It is well-known that the degree of third-order coherence of thermal light equals 6 \cite{zhou-2010,liu-2009}. Our measured degree of third-order coherence is much larger than 6, which confirms that three-photon superbunching is observed in our experiments based on the definition of superbunching \cite{ficek}. 

\begin{figure}[htbp]
\centering
\includegraphics[width=55mm]{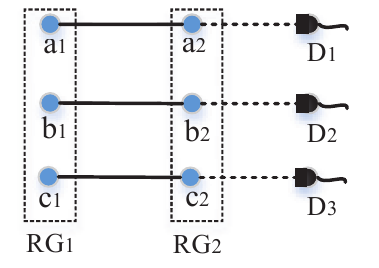}
\caption{Scheme for calculating the third-order coherence function of superbunching pseudothermal light. a$j$, b$j$, and c$j$ are the positions of photon a, b, and c on RG$_j$, respectively ($j=1$ and 2). }\label{al}
\end{figure}

Both quantum and classical theories can be employed to calculate the third-order coherence function of superbunching pseudothermal light \cite{sudarshan,glauber}. We will follow the method in Refs. \cite{zhou-2017, bai-2017} to calculate the third-order coherence function of superbunching pseudothermal light. There are 36 different alternatives for three photons to trigger a three-photon coincidence count in the scheme shown in Fig. \ref{al}, which is equivalent to the experimental setup shown in Fig. \ref{setup}. These 36 different alternatives are the product of two different groups of successive alternatives \cite{feynman}. One group corresponds to photons traveling from RG$_1$ to RG$_2$. The other group corresponds to photons traveling from RG$_2$ to three detectors. Let us take the first group for example to show how to list all the different alternatives. There are six different alternatives for photons at a1, b1, and c1 on RG$_1$ travels to a2, b2, and c2 on RG$_2$. For instance, one alternatives is photons at a1, b1, and c1 goes to a2, b2, and c2, respectively, which is shown by the black solid lines in Fig. \ref{al}. The corresponding three-photon probability amplitude can be written as $A_{a1a2}A_{b1b2}A_{c1c2}$. The other five probability amplitudes can be written in the same way as $A_{a1b2}A_{b1a2}A_{c1c2}$, $A_{a1c2}A_{b1b2}A_{c1a2}$, $A_{a1a2}A_{b1c2}A_{c1b2}$, $A_{a1b2}A_{b1c2}A_{c1a2}$,  and $A_{a1c2}A_{b1a2}A_{c1b2}$. The other group of alternatives for three photons at a2, b2, and c2 goes to D$_1$, D$_2$, and D$_3$ can be listed in the same way. If all the different alternatives are in principle indistinguishable, the third-order coherence function can be expressed as \cite{feynman}
\begin{eqnarray}\label{G2-1}
&&G^{(3)}(\vec{r}_1,t_1;\vec{r}_2,t_2; \vec{r}_3,t_3)=
\langle | (\sum_{j=1}^{6} A1_j)(\sum_{j=1}^{6}A2_j) |^2 \rangle,
\end{eqnarray}
in which $A1_j$ and $A2_j$ ($j=1$, 2, 3, 4, 5, and 6) are short for three-photon probability amplitudes in the first and second groups, respectively. $(\vec{r}_m,t_m)$ is the space-time coordinates of photon detection event at D$_m$ ($m=1$, 2, and 3). $\langle ... \rangle$ is ensemble average, which is equivalent to time average for ergodic system \cite{mandel-book}. 

The two groups of alternatives are independent in the scheme shown in Fig. \ref{al}. Equation (\ref{G2-1}) can be simplified as 
\begin{eqnarray}\label{G2-2}
&&G^{(3)}(\vec{r}_1,t_1;\vec{r}_2,t_2; \vec{r}_3,t_3)=
\langle | (\sum_{j=1}^{6} A1_j)|^2\rangle \langle |(\sum_{j=1}^{6}A2_j) |^2 \rangle,
\end{eqnarray}
where these two terms in the ensemble average correspond to the typical third-order coherence function of thermal light. From this point of view, the scheme in Figs. \ref{setup} and \ref{al} can be treated as a cascaded Hanbury Brown and Twiss interferometer \cite{HBT,HBT1}. If all the three detectors are in the symmetrical positions, which is the condition in our experiments, Eq. (\ref{G2-2}) becomes the third-order temporal coherence function by dropping the space coordinates. With the third-order temporal coherence function of thermal light given in Refs. \cite{zhou-2010,shih-book}, the third-order temporal coherence function in the scheme shown in Fig. \ref{setup} is
\begin{eqnarray}\label{G2-3}
&&G^{(3)}(t_1,t_2,t_3) \nonumber\\
&\propto& \prod_{l=1}^2[1+\text{sinc}^2\frac{\Delta \omega_l(t_1-t_2)}{2}+\text{sinc}^2\frac{\Delta \omega_l(t_2-t_3)}{2}\nonumber\\
&&+\text{sinc}^2\frac{\Delta \omega_l(t_3-t_1)}{2}\\
&&+2\text{sinc}\frac{\Delta \omega_l(t_1-t_2)}{2}\text{sinc}\frac{\Delta \omega_l(t_2-t_3)}{2}\text{sinc}\frac{\Delta \omega_l(t_3-t_1)}{2}]\nonumber,
\end{eqnarray}
where $\Delta \omega_l$ is the frequency bandwidth of the scattered light via RG$_l$ ($l=1$ and 2). Equation (\ref{G2-3}) can be easily generalized to $n$ rotating groundglasses by replacing the superscript 2 with $n$, where $n$ is a positive integer.

The observed results in Fig. \ref{super} can be fully interpreted by Eq. (\ref{G2-3}). For instance, the measured three-photon coincidence counts in Fig. \ref{super}(a) is proportional to Eq. (\ref{G2-3}). The maximal $CC$ in Fig. \ref{super}(a) is obtained when $t_1$, $t_2$, and $t_3$ are equal, which is consistent with the theoretical prediction in Eq. (\ref{G2-3}). The minimal $CC$ is obtained when the time difference,  $t_1-t_2$, $t_2-t_3$, and $t_3-t_1$, exceeds the coherence time of pseudothermal light. The ratio between the peak and constant background equals the degree of third-order coherence, $g^{(3)}(0)$, which is measured to be 22.55 in Fig. \ref{super}(c) and calculated to be 36 via Eq. (\ref{G2-3}). Figure \ref{super}(b) is the section of the measured third-order temporal coherence function in Fig. \ref{super}(a) when $t_1$ equals $t_3$. The red curve in Fig. \ref{super}(b) is the theoretical fitting of the data by assuming $t_1$ equals $t_3$ in Eq. (\ref{G2-3}),
\begin{eqnarray}\label{G2-31}
&&G^{(3)}(t_1-t_2) \propto \prod_{l=1}^2[2+4\text{sinc}^2\frac{\Delta \omega_l(t_1-t_2)}{2}].
\end{eqnarray}
The measured ratio between the peak and background in Fig. \ref{super}(b) is 7.72, which is less than the theoretical value, 9. The measured ratio does not equal $g^{(3)}(0)$ due to the background is obtained when $t_1$ equals $t_3$. Similar conclusions hold for Fig. \ref{super}(d), in which it is a section of Fig. \ref{super}(a) when $t_2$ equals $t_3$. The measured  ratio between the peak and background is 7.10. Figure \ref{super}(c) is the section of the measured third-order coherence function when $t_1-t_2$ equals $t_2-t_3$, in which Eq. (\ref{G2-3}) is simplified as
\begin{eqnarray}\label{G2-32}
&&G^{(3)}(t_1-t_2) \nonumber\\
&\propto& \prod_{l=1}^2[1+2\text{sinc}^2\frac{\Delta \omega_l(t_1-t_2)}{2}+\text{sinc}^2{\Delta \omega_l(t_1-t_2)}\\
&&+2\text{sinc}^2\frac{\Delta \omega_l(t_1-t_2)}{2}\text{sinc}{\Delta \omega_l(t_1-t_2)}]\nonumber,
\end{eqnarray}
The constant background is obtained when $t_1-t_2$ exceeds the coherence time of pseudothermal light. $\Delta \omega_l$ is assumed to be identical for pseudothermal light generated by RG$_1$ and RG$_2$ in the above fitting processes. The ratio between the peak and background is measured to be 22.55, which is less than the theoretical value, 36. It is the measured degree of third-order coherence of superbunching pseudothermal light.

It is concluded that the visibility of ghost imaging with thermal light is positively correlated with the degree of optical coherence of thermal light \cite{gatti-2004,gatti-jmo}. The measured $g^{(3)}(0)$ equals 22.55, which is much larger than the theoretical maximal $g^{(3)}(0)$ of thermal light, 6 \cite{liu-2009,shih-book}. Even through when two of these three photon detection events are in the same space-time coordinate, the measured ratio between the peak and background is also larger than 6, which can also be employed to increase the visibility of ghost imaging. We have obtained  $g^{(3)}(0)$ equaling 22.55 with only two RGs. $g^{(3)}(0)$ can be increased to $(3!)^n$ by employing $n$ RGs \cite{bai-2017}. For instance, $g^{(3)}(0)$ can reach 216 ($(3!)^3$) with three RGs. With four RGs, $g^{(3)}(0)$ can be as large as 1296 ($(3!)^4$). Even through the measured $g^{(3)}(0)$ is less than the theoretical value, the visibility of ghost imaging with superbunching pseudothermal light can be much larger than the one of ghost imaging with thermal or pseudothermal light.

As mentioned before, the premise to have Eq. (\ref{G2-1}) is that all the 36 different alternatives to trigger a three-photon coincidence count are in principle indistinguishable, which is the key to observe three-photon superbunching in our experiments. It is implemented by putting a pinhole before RG$_l$. The size of the pinhole is less than the spatial coherence area of the scattered light by RG$_{l-1}$ in the pinhole plane. Since photons within a coherence volume are in principle indistinguishable \cite{mandel-book}, it is impossible to distinguish these different ways to trigger a three-photon coincidence count in the scheme shown in Fig. \ref{setup} when the size of the pinhole is smaller than the size of spatial coherence area of the scattered light. 

In summary, we have experimentally observed three-photon superbunching with superbunching pseudothermal light, which is  generated by single-mode continuous-wave laser light and linear optical elements such as pinholes, lens, and rotating groundglasses. The degree of third-order coherence of superbunching pseudothermal light is measured to be 22.55 with two rotating groundglasses, which is much larger than the one of thermal or pseudothermal light. It is predicted that the degree of third-order coherence of superbunching pseudothermal light can further be increased by adding more rotating groundglasses. For instance, the degree of third-order coherence can be as large as 216 and 1296 for three and four rotating groundglasses, respectively. From the quantum optics point of view, the key to observe three-photon superbunching is that all the different alternatives to trigger a three-photon coincidence count detection event are in principle indistinguishable. The constructive-destructive three-photon interference, \textit{i.e.}, the superposition of three-photon probability amplitudes, are the reason for three-photon superbunching in the superbunching pseudothermal light. The results in this letter is helpful to understand the physics of third-order interference of light and increase the visibility of thermal light ghost imaging.

\textbf{Funding.} National Natural Science Foundation of China (NSFC) (11404255), 111 Project of China (Grant No.B14040), and Fundamental Research Funds for the Central Universities.

\end{document}